\documentclass[appendixfloats,tighten,iop,numberedappendix]{emulateapj}

\begin{document}
\title{The Case for Standard Irradiated Accretion Disks in Active Galactic Nuclei}
\author{Doron Chelouche\altaffilmark{1}}
\altaffiltext{1} {Department of Physics, Faculty of Natural Sciences, University of Haifa, Haifa 31905, Israel; doron@sci.haifa.ac.il}
\shortauthors{Chelouche}
\shorttitle{The Case for Standard Accretion Disks in AGN}

\begin{abstract}

We analyze the broadband photometric light curves of Seyfert 1 galaxies from the \citet{ser05} sample and find that a) perturbations propagating across the continuum emitting region are a general phenomenon securely detected in most cases, b) it is possible to obtain reliable time-delays between continuum emission in different wavebands, which are not biased by the contribution of broad emission lines to the signal, and that c) such lags are consistent with the predictions of standard irradiated accretion disk models, given the optical luminosity of the sources. These findings provide new and independent support for standard accretion disks being responsible for the bulk of the (rest) optical emission in low-luminosity active galactic nuclei (AGN). We interpret our lag measurements in individual objects within the framework of this model and estimate the typical mass accretion rate to be $\lesssim 0.1\,{\rm M_\odot~yr^{-1}}$, with little dependence on the black hole mass. Assuming bolometric corrections typical of type-I sources, we find tentative evidence for the radiative efficiency of accretion flows being a rising function of the black hole mass. With upcoming surveys that will regularly monitor the sky, we may be able to better quantify possible departures from standard self-similar models, and identify other modes of accretion in AGN.

\end{abstract}

\keywords{
Accretion, accretion disks ---
galaxies: active ---
methods: data analysis ---
quasars: general ---
techniques: photometric
}

\section{Introduction}

The current physical paradigm of accretion onto black holes (BHs) originates in the works of \citet{sal64}, \citet{zel64}, \citet{lyn69}, \citet{sh72}, \citet{nt73}, and \citet{ss73}. In the standard model of accretion disks, matter gradually accretes onto the BH while losing angular momentum due to viscous shear. As the accreted gas becomes more tightly bound to the BH, its gravitational energy is locally dissipated and emitted as local black-body radiation. The combined emission from a range of disk annuli around BHs typical of active galactic nuclei (AGN) gives rise to a powerlaw emission in the optical band. The radiative efficiency of accretion processes through a thin disk depends largely on the gas ability to radiate close to the innermost stable orbit, which depends on the BH spin. Under favorable conditions, the total energy extracted is $\lesssim 30$\% of the gas rest energy \citep{nt73,nob11}, and most of it is emitted at extreme-UV (EUV) energies for the parameter range typical of AGN. 

We note that aside from thin accretion disks, a variety of additional accretion modes likely exist, such as advection dominated accretion flows \citep{ny94}, and slim disk solutions \citep{ab88}. Other configurations, such as ADIOS \citep{bb99} and CDAF \citep{qg00}, where only a small fraction of the gas ultimately gets accreted by the BH, are also possible.

Qualitatively, the rapid time-varying flux level from AGN is consistent with a compact source, which, by causality arguments, implies a BH-scale phenomenon. In particular, geometrically-thin accretion disk models have been very successful in explaining the prominent continuum emission of AGN over a broad range of wavelengths, from the optical to the soft X-ray bands. Quantitatively, however, very little is known with confidence about the physical properties of accretion flows, and many uncertainties remain about their nature. For example, the (wavelength-dependent) size of the continuum emitting region is rather loosely constrained by causality arguments, and some tension has recently risen between model predictions and microlensing-assisted measurements of the optically-emitting region in a few lensed quasars \citep[but see also \citet{poi08}]{bla11}.  Whether this tension is simply due to measurement systematics or whether it points to new, previously unrealized physics of accretion flows \citep{da11}, is currently unclear. Providing reliable constraints on accretion disk models is therefore of considerable importance for shedding light on processes of gas accretion onto compact objects. 

Reverberation mapping is a widely-used technique for probing spatially unresolved regions in astrophysical systems. This method, as applied to spectroscopic light curves of AGN, proved to be essential for studying the broad emission line region (BLR) in those objects, and for determining the masses of the BHs powering them. Similarly, by searching for, and studying, reverberating continuum signals over a range of wavelengths, one may be able to deduce the scale of accretion flow, thereby contrasting accretion disk model predictions with direct observables other than the shape of the spectral energy distribution \citep{mal83,ln89}. Indeed, \citet{col98,col01b} carried out the only intensive spectroscopic campaigns of two Seyfert 1 galaxies (NGC\,7469, and the narrow line object Akn\,564) leading to wavelength-dependent time-delay maps, which are in qualitative agreement with thin accretion disk models. However, being the only objects for which such studies has been carried out to date (mainly due to the imposing requirements on high signal-to-noise ratios, S/N, and high-cadence observations), it remains to be established that this phenomena is general, and to understand its manifestation in AGN of different properties, such as BH mass and luminosity. 

While several attempts have been made to use multi-band photometric data to constrain the physics of accretion processes in AGN \citep{ser05,bac09,cak07}, the scatter in the measured inter-band time-delays was large and, as discussed in \citet{cz12}, results were biased by the finite contribution of the broad emission lines to the signal. In particular, \citet{cz12} showed that the determination of continuum time-delays and of line-to-continuum time-delays must be done self-consistently when broadband photometric data are concerned. In this work we apply their formalism to carry out reliable reverberation mapping analysis of the accretion disks in a sample of low-luminosity AGN (Seyfert 1 galaxies) from \citet{ser05}. 

This paper is organized as follows:  the AGN sample is presented in \S2, and the method of analysis briefly outlined in \S3. The results are presented in \S4, with their interpretation within the framework of irradiated thin accretion disk models given in \S5. Section 6 discusses our findings in the context of AGN physics and the prospects for reverberation mapping of accretion disks using upcoming photometric surveys. The summary follows in \S7.

\begin{figure}
\plotone{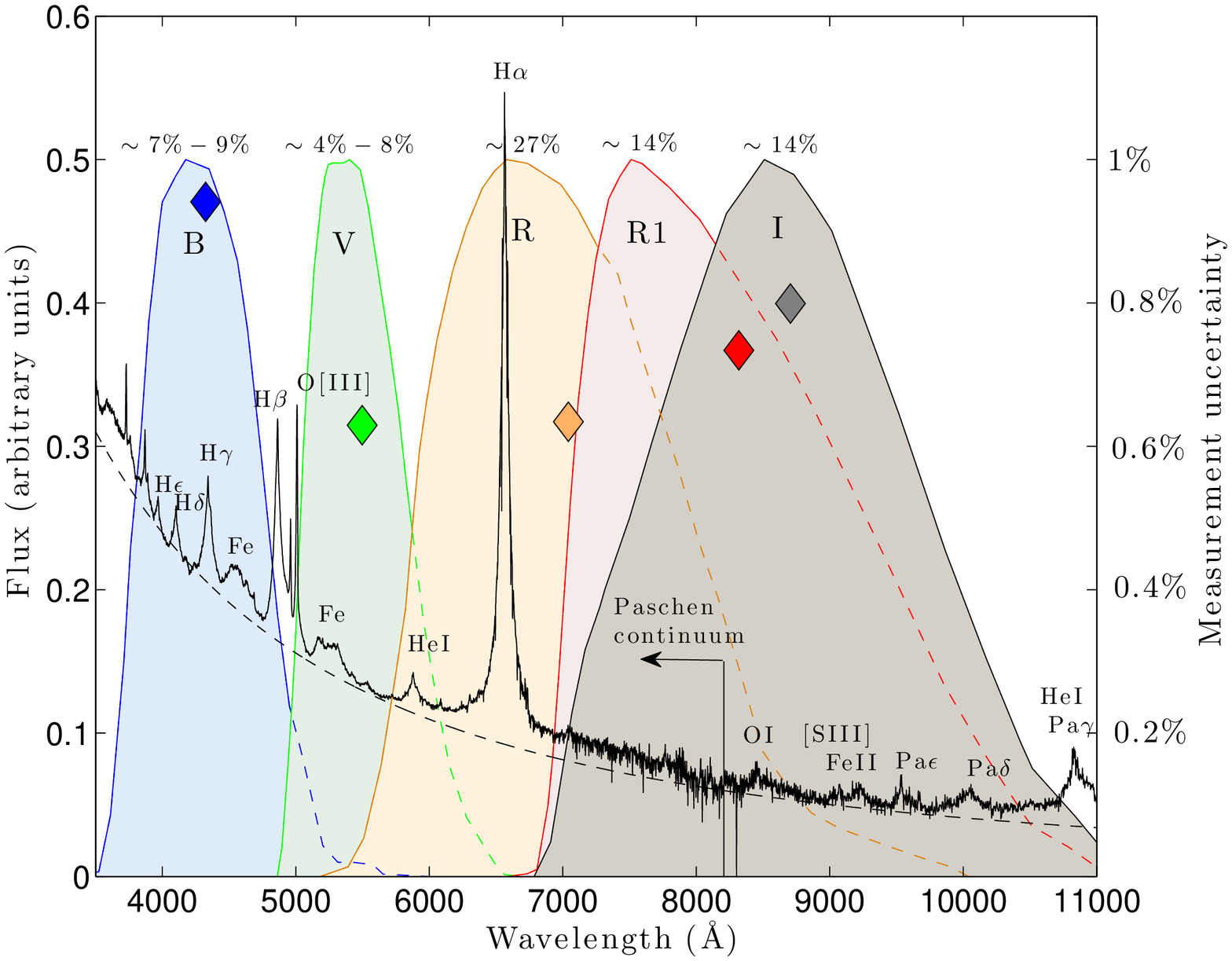}
\caption{Filter throughput curves for the Crimean telescope \citep{ser05} overlaid with the $z=0$ composite AGN spectrum of \citet{gl06}. Also shown are the typical measurement uncertainties and the throughput-weighted central wavelengths for each band (colored diamonds). The percentages denote the relative contributions of all emission components other than the powerlaw (dashed line) to the composite spectrum in each band (lower values correspond to a case where iron emission is ignored; see the text).}
\label{filters}
\end{figure}

\section{Sample}

Here we restrict our analysis to objects in the \citet{ser05} sample of low-$z$ Seyfert 1 galaxies for which the BLR size, hence also the BH mass, $M_{\rm BH}$, have been independently constrained by means of spectroscopic reverberation mapping \citep{ben09,gr12,pet04}. The reason for that will be further clarified below. Some properties of the sample are given in table 1 [see also \citet{ser05} and \citet{cz12} for more details]. 

We estimate the monochromatic optical luminosity at 5100\AA\, $L_{\rm opt}\equiv \lambda L_\lambda (5100{\rm \AA})$ for objects in the \citet{ser05} sample in several ways to bracket the relevant luminosity range under a plausible range of underlying assumptions. We note, however, that by using different datasets to estimate the AGN luminosity, some scatter in the luminosity may be expected on grounds of object-variability as well as by the different starlight contribution admitted by the apertures used by different studies. 

We first consider the results of \citet{cak07} who quote $V$-band fluxes (we take the mean between their reported high- and low-states), and convert those to luminosity assuming concordance cosmology, while taking into account their reported Galactic and intrinsic reddening values (derived from the Balmer line ratio), and assuming $R_V=3.1$. We also correct for the contribution of the host using their best-fit values. As an alternative estimate for the optical luminosity, we consider the results of  \citet{ben09} who were able to directly subtract the host's contribution from the signal, but did not correct for intrinsic extinction. Unsurprisingly, their reported luminosities tend to be the lowest of all other estimates yet still larger than those reported by \citet{ben06}. As our third estimate, we take the absolute $B$-band magnitudes reported by \citet{ser05}, and convert those to luminosities assuming the conversion factors appropriate for low-$z$ quasars in \citet{kas00}. Doing so we effectively neglect reddening effects as well as potential contribution from the host galaxy to the $B$-band flux. Lastly, we consider the luminosities previously reported by \citet{kas00,pet04} for some objects in our sample. The optical luminosities reported in Table 1 are the geometric means of all estimates. We note the especially large range of luminosities reported in the literature for Mrk\,6, NGC\,4151, and NGC\,3516 \citep{ben09,cak07}.

In this work we use the published $B,~V,~R,~R1$ and $I$ light curves from \citet{ser05} and note that, at the redshift of the sample, the $B$-band includes high-order Balmer line emission, and the $V$-band is relatively devoid of strong emission lines (Fig. \ref{filters}). The $R$-band contains the prominent H$\alpha$ emission line, with redder bands showing little contribution from strong emission lines. The best S/N for objects in this sub-sample is obtained for the $V$- and $R$-bands.

\section{Method}

We use the multivariate correlation function (MCF) method of \citet{cz12}, which is able, under favorable conditions, to separate line-to-continuum  and continuum-to-continuum time-delays ($\tau_l$ and $\tau_c$, respectively) from the broadband photometric light curves of quasars. The reader is referred to \citet{cz12} for a complete description of the technique, and its implementation for a portion of the \citet{ser05} dataset. In a nutshell, this technique eliminates the biases inherent to simple cross-correlation analyses that are encountered when several reverberating emission components, such as continuum and emission lines, contribute to the signal. Specifically, the algorithm seeks the best agreement between the (fluxed) light curve in one band,  $f_X(t)$, and its model, $f_X^m(t)$, as a function of the model parameters. More specifically, $f_X(t)$, is assumed to include at most two prominent echoed signals, with either positive or negative time-delays, of a light curve in some other band, $f_V(t)$. An adequate first order approximation for $f_X(t)$ is then given by $f_X^m(t;f_V,\alpha,\tau_c,\tau_l)\equiv (1-\alpha)f_V(t-\tau_c)+\alpha f_V(t-\tau_l)$, where $\alpha$ determines the relative contribution of the two signals to the broadband flux. Specifically, a parameter combination is sought in three dimensional phase space, which maximizes the (linearly-interpolated) Pearson correlation coefficient between the model and the data, $R(f_X,f_X^m)$. As demonstrated in \citet{cz12}, the use of priors on one or more parameter values, when available, may lead to improved constraints on the remaining parameters of the problem.

As discussed in \S2 (see also Fig. \ref{filters}), the $V$-band is relatively devoid of broad emission line contribution to its flux, and its light curve is typically characterized by a high S/N data. Specifically, a single powerlaw fit to the composite spectrum of \citet{gl06} implies that the contribution of non-powerlaw continuum emission to this band is on the order of 5\% or less, hence lower than that for the other bands (Fig. \ref{filters}).  For these reasons, we consider the $V$-band light curves as our "pure" continuum light curve, hence $f_V(t)$ in the above notation, and use it to construct $f^m_X(t)$ where $X\in[B,R,R1,I]$. Upon maximizing $R(f_X,f_X^m(f_V,\alpha,\tau_c,\tau_l))$, we find $\tau_c,\tau_l$, and $\alpha$ for each of the bands.

\section{Analysis}

As discussed in \citet{cz12}, determining $\tau_c$, which is a measure of the accretion disk size (see below), requires, in principle, the simultaneous determination of $\tau_l$ and $\alpha$. We first apply the full reverberation mapping analysis for a sub-sample of objects from the \citet{ser05}, and then, motivated by its results, extend the analysis to the full sample using spectroscopic priors on $\tau_l$.

\begin{table}
\begin{center}
\caption{The sample of AGN}
\begin{tabular}{llccll}
ID & $z$  & $L_{\rm opt}$ &  $M_{\rm BH}$ & $\tau_c^0$ & $\dot{\mathcal{M}}$  \\
(1)  & (2) & (3) & (4) & (5) & (6) \\
\tableline
3C\,390.3 	& 0.056 &   43.9	&	8.46 	& $1.2^{+0.4}_{-0.2}(1.5)$ &$0.02$   \\
Akn\,120		& 0.033 &    44.1 	&	8.18 & $3.0^{+0.1}_{-0.5}(1.7)$ &$0.4$    \\
Mrk\,335 		& 0.026 &   43.8 	&	7.15 & $2.14^{+0.20}_{-0.14}(2.10)$ &$2.8$          \\
Mrk\,509 		& 0.035 &   44.2	&	8.16 & $8.9^{+0.3}_{-0.9}(6)$ &$13$               \\
Mrk\,6 		& 0.019 &    43.4 	&	8.13 & $0.9^{+0.2}_{-0.2}(1.3)$ &$0.03$   \\
Mrk\,79 		& 0.022 &    43.7 	&	7.72 & $0.78^{+0.08}_{-0.04}(1.0)$ &$0.03$   \\
NGC\,3227 	& 0.004 &   42.4 	&	7.63 & $0.58^{+0.04}_{-0.14}(0.20)$ &$0.01$    \\
NGC\,3516 	& 0.009 &    43.1 	&	7.49 & $0.90^{+0.16}_{-0.18}(1.0)$ &$0.07$          \\
NGC\,4051 	& 0.002 &    41.9 	&	6.28 & $0.22^{+0.08}_{-0.10}(0.3)$ &$0.02$    \\
NGC\,4151 	& 0.003 &    42.6 	&	6.66 & $0.50\pm0.02(0.53)$ &$0.07$   \\
NGC\,5548 	& 0.017 &    43.3 	&	7.83 & - & -                                           \\
NGC\,7469 	& 0.016 &   43.6 	&	7.09 & $ 0.72^{+0.10}_{-0.04}(1.0)$  &$0.09$     \\
\tableline
\end{tabular}
\end{center}
Column description: (1) Object name, (2) redshift, (3) optical luminosity at 5100\AA\ in Log\,(${\rm erg~s^{-1}}$) units (see text), (4) mass in Log\,(${\rm M_\odot}$), (5) time-lag in days where the best-fit model is forced to go through zero lag at 5500\AA\ (values in parenthesis correspond to fits that were not forced to go through zero lag, i.e., the dashed line models in Fig. \ref{tiles}) , (6) effective mass accretion rate in units of ${\rm M_\odot~yr^{-1}}$.  
\end{table}

\subsection{A sub-sample of three AGN from \citet{ser05}}

We restrict the analysis in this section to objects in the \citet{ser05} sample, which gave high confidence-level ($P>90\%$) results in the \citet{cz12} analysis of the $V$ and $R$ data. For those objects, solutions for $\tau_c,~\tau_l$ and $\alpha$ were most secure, and $\tau_l$ values showed good agreement with independent constraints from spectroscopic reverberation mapping studies \citep[and references therein]{pet04,gr12}. Our sub-sample of objects therefore includes: NGC\,5548, NGC\,4151, and NGC\,3516 \citep[see their Table 1 for the relevant data]{cz12}. De-trending of the light curves by a first degree polynomial was used throughout.

\begin{figure}
\epsscale{1}
\plotone{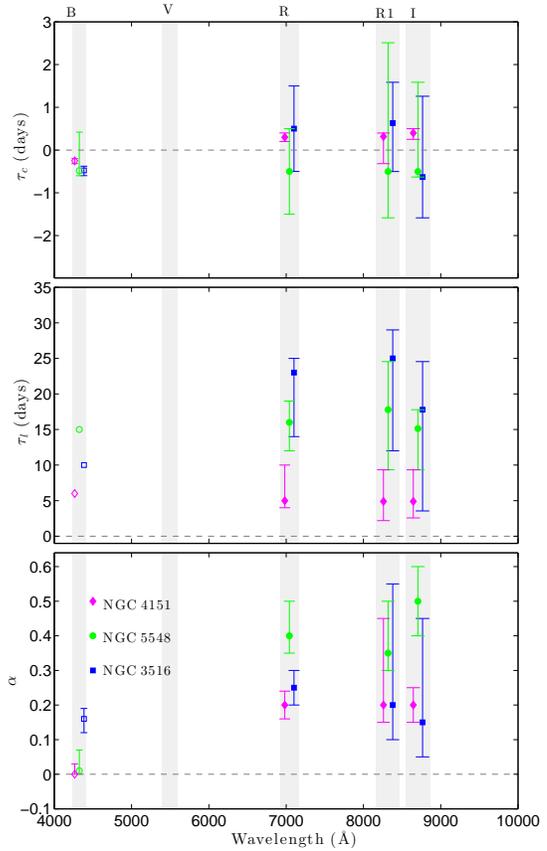}
\caption{The continuum-to-continuum ($\tau_c$; upper panel), line-to-continuum ($\tau_l$; middle panel), and relative BLR emission contribution to the band ($\alpha$; lower panel) relative to the $V$-band light curve for three AGN (see legend). Empty symbols correspond to spectroscopically-determined Balmer line lags. Clearly, $\tau_l$ values are comparable in all bands, and consistent with the spectroscopic values (see text). A trend wherein $\tau_c$ increases with wavelength is seen for NGC\,4151 and NGC\.3516. Deduced values for $\alpha$ are consistent with naive spectral decomposition (Fig. \ref{filters}).  Small wavelength shifts between objects were applied for clarity.}
\label{delays}
\end{figure}

Solutions for $\tau_c$ and $\tau_l$, and $\alpha$ are shown in figure \ref{delays} for the various bands. The quoted values reflect on the mode of the distribution functions obtained from Monte Carlo simulations that introduce Gaussian measurement noise in accordance with the reported measurement uncertainties \citep{ser05}. Reported error bars reflect on $\pm68\%$ percentiles around the mode value \citep{cz12}. 

Statistically significant $\tau_l$-values are detected in all three objects and in all bands but for the noisier $B$ band (not shown). In particular, for a given object, the deduced values are consistent for the different bands, as may be expected if emission from similar locations in the BLR  contributes to their flux. As noted in \citet[see their Table 1]{cz12}, these lags are also consistent with the spectroscopically measured Balmer time-delays. This is indeed expected given that hydrogen emission dominates the non-powerlaw emission in those bands either through Balmer line emission, or via Paschen lines and continuum emission (Fig. \ref{filters}). 

As far as results for the $B$-band are concerned, we note that higher-order Balmer line and iron blend emission are important contributors to the flux, and while the former originate from regions comparable to the H$\alpha$ and H$\beta$ emission regions \citep{ben10}, the location of the latter is less secure with some studies placing it on scales comparable or slightly larger than the Balmer line emission region \citep{bi10,hu08}; see also the new results of \citet{bar13,raf13} who find that the region is comparable in size to that which emits the H$\alpha$ line \citep{ben10}. For these reasons, in our analysis of the $B$-band data, we henceforth {\it set} $\tau_l$ to the spectroscopically-measured Balmer line time-delay \citep[see their table 1 and references therein]{cz12}, and determine $\tau_c$ and $\alpha$ for the $B$-band under this prior (see \S4.1.1).

The deduced values for $\alpha$ for the redder bands are of order $20\%$ (NGC\,5548 results in $\alpha\sim 40\%$; see Fig. \ref{delays}). In comparison, the $B$-band data implies a smaller (5\%-10\%) relative contribution of a lagging emission component to the band. Indeed, a qualitative spectral decomposition of the \citet{gl06} composite spectrum suggests that the contribution of non-powerlaw emission to the bands is of the same order\footnote{It should be borne in mind that $\alpha$ measures the relative contribution of a non-powerlaw (lagging) emission component to the variable part of the light curve. In particular, in the presence of a constant additive flux to the redder bands (e.g., host galaxy contribution), the estimated $\alpha$ from the mean quasar template could be lower than its actual value.} (Fig. \ref{filters}). 

Our $\tau_c$ measurements relative to the $V$-band show a qualitative trend whereby $\tau_c$ increases with wavelength from negative values for the $B$-band to positive lags for the redder bands. This is most notable for NGC\,4151 and less so for NGC\,3516 due to the larger uncertainties. Results for NGC\,5548, while formally negative, are consistent with a zero lag in all bands, with only the upper uncertainty envelope hinting on a possible increase of $\tau_c$ with wavelength.

\begin{figure}
\plotone{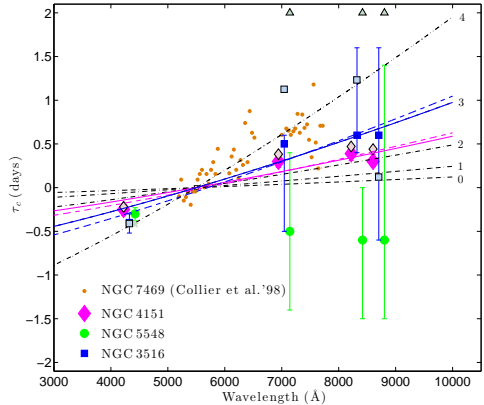}
\epsscale{1.0}
\caption{Solutions for $\tau_c$ using a spectroscopic prior on $\tau_l$ for three AGN (see legend and compare to the upper-panel of Fig. \ref{delays}). Also shown, for comparison, are the spectroscopic results for NGC\,7469 by \citet{col98}. Over-plotted are $\tau_c =0.1 \times 2^n [(\lambda/5500\,{\rm \AA})^{4/3}-1]$ models with $n$ indicated next to each curve.  Best-fit models, in the least squares sense, appear as colored curves, where the fit is forced to go through $\tau_c(5500\,{\rm \AA})=0$\,days (solid lines), and when it is not (colored dashed lines). Light shaded symbols denote the interband time-delays of  \citet[those for NGC\,5548 are $\gtrsim 5$\,days and are marked here as lower limits]{cak07}. Small wavelength shifts were applied for clarity.} 
\label{tc}
\end{figure}

\begin{figure}
\plotone{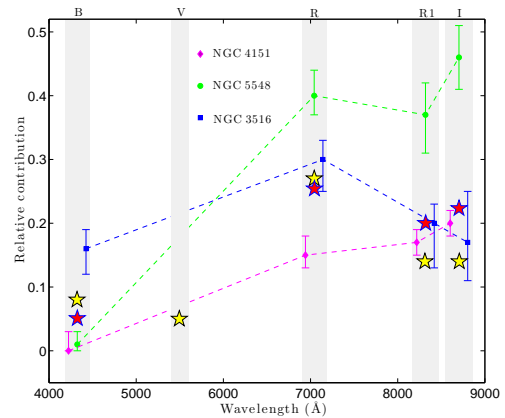}
\caption{Solutions for $\alpha$ using a spectroscopic prior on $\tau_l$ for three AGN (see legend and compare to the bottom-panel of Fig. \ref{delays}). Also shown are $\alpha$-estimates based on spectral decomposition of the \citet{gl06} composite  (yellow pentagrams). The median $\alpha$-values for 12 objects from the \citet{ser05} sample, are shown as red pentagrams. Small wavelength shifts were applied for clarity.}
\label{al}
\end{figure}

\subsubsection{Setting priors on $\tau_l$}

The consistency between the $\tau_l$-values deduced for the different bands, the agreement between those and the spectroscopic time-lag measurements for the Balmer emitting region (Fig. \ref{delays} and \citet{cz12}), as well as the reasonable results for $\tau_c$ and $\alpha$ obtained for the $B$-band when using priors, motivate us to consider a refined analysis approach, where the number of degrees of freedom is reduced. In particular, we henceforth {\it set} $\tau_l$ to the spectroscopically-measured Balmer emission line lag \citep[see their Table 1 for the adopted values, and note that we neglect the measurement uncertainty on $\tau_l$, as well as the weak inverse trend between the time-delay and the increasing line order \citep{ben10}]{cz12}, and constrain the values for $\tau_c$ and $\alpha$ under this prior in the remaining two-dimensional phase space for all bands. The results are shown in figures \ref{tc} and \ref{al} for the three objects in our sub-sample. 

Generally, the results obtained by setting priors on $\tau_l$ are consistent with those deduced without priors, yet the uncertainties are reduced, and higher significance $\tau_c$ measurements are obtained. Specifically, the deduced $\tau_c$ values  are in qualitative agreement with irradiated accretion disk model predictions, where the lag increases with wavelength as one views larger emitting regions.  By and large, the deduced $\tau_c$-values are consistently below those identified by inter-band time-delays \citep[most notably for NGC\,5548; see our Fig. \ref{tc}]{cak07,ser05}, which are contaminated by emission lines signal \citep{cz12}, and are in better agreement with those obtained for a similar luminosity object using spectroscopic means \citep[see also \S4.2]{col98}

The solutions for $\alpha$ are now better-defined and show fair agreement with qualitative spectral decomposition estimates (Fig. \ref{al} and more in \S4.2). For example, NGC\,3516 shows a maximal $\alpha$-value in the $R$ band, which can be attributed to the dominant contribution of H$\alpha$ (Fig. \ref{filters}). Results for NGC\,4151 imply an overall reduced contribution of emission lines to the flux (more accurately, flux variance) compared to NGC\,3516, but that which monotonically rises toward the redder parts of the spectrum. There are various possible reasons for that, among which a BLR that varies less on the relevant timescales, and/or continuum emission that varies less in the redder parts of the spectrum, or is suppressed altogether due to disk truncation (it is not possible to robustly test either of these scenarios using the present dataset). Lastly we note the case of NGC\,5548 that exhibits $\alpha$-values that are in excess of what is expected from spectral decomposition, and in addition leads for formally negative (yet consistent with zero) lags for the redder bands. Together, these facts lead us to doubt the validity of the $\tau_c$ and $\alpha$ solution in this case, and we exclude this object from further analysis.

\begin{figure}
\epsscale{1}
\plotone{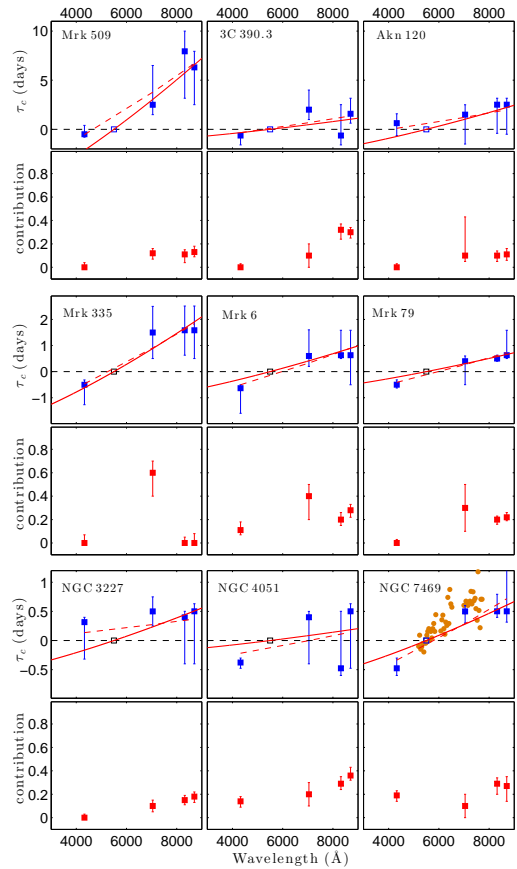}
\caption{Results for $\tau_c$ (top panels in each group of two rows; note the different ordinate scales in each group) and $\alpha$ (bottom panels in each group) using a spectroscopic prior on $\tau_l$. Note the consistency with the spectroscopic $\tau_c$-measurements of \citet[orange dots]{col98} for NGC\,7469. In all objects but for NGC\,3227 and Akn\,120, a monotonic $\tau_c(\lambda)$ behavior is observed, as theoretically expected. Over-plotted in red curves are model fits to the data, where solid lines are forced to go through $\tau_c(5500\,{\rm \AA})=0$\,days (Eq. \ref{tce}). In $\sim 50$\% of the objects, $\alpha$ peaks in the $R$-band due the relatively prominent contribution of the H$\alpha$ line to the flux (see text). Finite $\alpha$ is obtained also for the reddest bands, and is  due to the relatively large contribution of Paschen line and continuum emission to the {\it variable} component of the flux at those wavelengths.}
\label{tiles}
\end{figure}

\begin{figure}
\epsscale{1.0}
\plotone{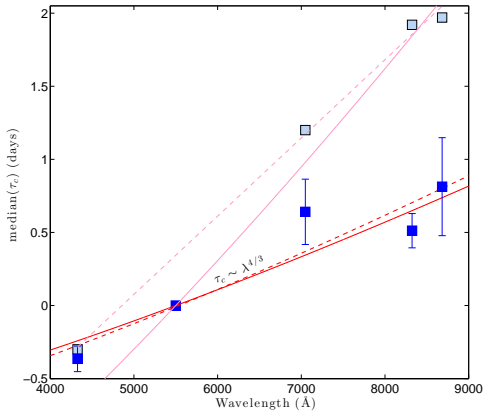}
\caption{The median $\tau_c(\lambda)$ relation for all objects, excluding NGC\,5548, shows consistency with irradiated accretion disk models' predictions (dashed and solid lines, with the latter being forced to go through zero lag at 5500\AA; see text). Error bars were estimated using Monte Carlo simulations that consider the measurement uncertainty on $\tau_c(\lambda)$. Also shown, for comparison, are the median interband time-delays from \citet[light shaded points]{cak07} that lead to a factor $\sim 2$ steeper slope (light-red curves) hence over-estimate $\dot{\mathcal{M}}$ by a factor $\sim 8$ (see the text).}
\label{med}
\end{figure}

\subsection{The full \citet{ser05} sample}

We next analyze the remaining 9 objects in the  \citet{ser05} sample using spectroscopic priors on $\tau_l$ (Table 1 in \citet{cz12}). Results for most objects are consistent with $\tau_c$ rising monotonically with wavelength from negative values for the $B$-band to positive values for the redder bands (Fig. \ref{tiles}). In particular, results for NGC\,7469 are in good agreement with the spectroscopic lag measurements of \citet{col98}. NGC\,3227 and Akn\,120 are exceptions as $\tau_c$ for the $B$-band is marginally positive, in contrast to naive model expectations where the bluer bands should lead the redder ones. The reason for that is currently unclear, and may have to do with the particular light curve behavior observed, sampling, or residual signal from emission lines. Results for the $R1$- and $I$-bands are largely consistent as the centroid wavelengths of both bands are quite similar\footnote{Note that the wavelength which actually characterizes the bands is a (filter) response-weighted average over the light curve variance at each wavelength. This, however, is not known, and centroid values, assuming a flat quasar spectrum, are used instead.} (Fig. \ref{filters}).

Considering the statistical properties of the entire sample (Fig. \ref{med}), we note that the median $\tau_c$ is consistent with being monotonically rising with wavelength, i.e., with geometrically-thin accretion disk model predictions where $\tau_c(\lambda) \sim \lambda^{4/3}$ \citep{col98}\footnote{Note, however, that the limited wavelength range probed here does not allow us to provide physically-meaningful constraints on the powerlaw index.}. In comparison, the best-fit curve obtained using the median results of \citet{cak07} is much steeper, with an effective linear slope that is $\sim 2$ times greater than the one found here. As discussed above, this is likely because of BLR emission contamination of the signal in cross-correlation analyses \citep{cz12}.

Turning our attention to $\alpha$, we note that it is, typically, smallest for the $B$-band, peaks in the $R$ band, and somewhat declines, yet remains finite also for the $R1$- and $I$-bands (Fig. \ref{tiles}). This behavior is expected from a naive spectral decomposition of the \citet{gl06} composite into powerlaw and non-powerlaw emission components (Fig. \ref{al}). Particular objects (e.g., NGC\,3227) show, however, a more monotonic increase of $\alpha$ with wavelength, as discussed in \S4.1.1. Further investigation into the different $\alpha$ behavior observed in different objects is outside the scope of this work.

\section{The case for standard accretion disks}

Motivated by the fact that $\tau_c(\lambda)$ behavior is consistent, at least on average, with the model expectations (Fig. \ref{med}), and that reasonable values are obtained for both the continuum lags as well as for the relative contribution of line and continuum processes, for all bands, we shall now interpret our findings for individual objects within the framework of the irradiated geometrically-thin accretion disk model \citep[see also the Appendix]{col98,cak07}. Specifically, we turn to figures \ref{tc} and \ref{tiles}, where the measured $\tau_c(\lambda)$ values are fitted, in the least squares sense\footnote{We treat the time-delays associated with each band as independent measurements yet note that this is not strictly true since the $V$-band light curve is used to determine each of the lags. Further, correlated errors between the light curves in different bands could be an issue, although the MCF is less sensitive to those \citep{cz12}.}, by a model for a self-similar geometrically-thin accretion disk that is illuminated by a point source\footnote{This scenario is consistent with the rapid variability timescales observed in AGN with respect to the dynamical times in the optically emitting region. With that being said, this assumption is bound to break down close enough to the black hole, i.e., short enough emission wavelengths.}, for which \citep{col98,cak07}
\begin{equation}
\tau_c(\lambda_X)=\tau_c^0 \left [ \left ( \frac{\lambda_X}{5500\,{\rm \AA}} \right )^{4/3}-1 \right ],
\label{tce}
\end{equation}
where $\lambda_X$ is the centroid wavelength of band $X$ obtained by assuming a flat quasar continuum and the filter's throughput curves (Fig. \ref{filters}). For example, we find, $\tau_c^0=0.5\pm0.02$\,days for NGC\,4151, and $\tau_c^0=0.90^{+0.16}_{-0.18}$\,days for NGC\,3516\footnote{Here, $\pm34$\% percentiles around the mode value are reported, which are adequate to bracket the solution given the simple, single-peaked, functional behavior of the MCF is the constrained phase space \citep{cz12}.}. For comparison, we also experimented with less physical models, which were not restricted to pass through $\tau_c(5500\,{\rm \AA})=0$\,days (effectively it allows for deviations of the centroid broadband wavelengths from the assumed values), and find consistent results for most objects (see Table 1). Objects for which significant differences are encountered between the two fitting methods include Mrk\,509, Akn\,120, and NGC\,3227, with the latter two having their blue continua formally leading the $V$-band, yet still consistent with a zero lag.

\subsection{The Lag-Luminosity Relation}

\begin{figure}
\plotone{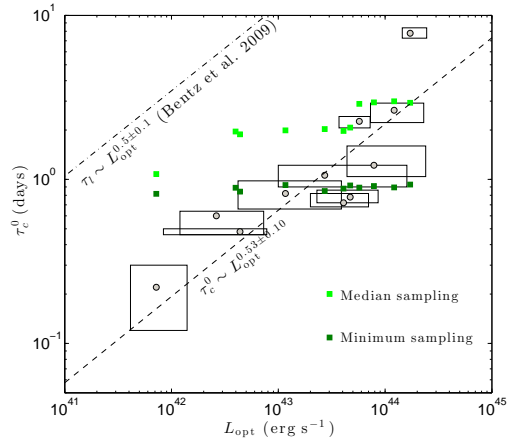}
\caption{The continuum lag-luminosity relation overlaid with the best fit model ($\tau_c \propto L_{\rm opt}^{0.53\pm0.10}$), which takes into account measurement uncertainties on the lag but not on the luminosity. For comparison we show the median (light green) and the minimal (dark green) sampling periods for our sources. Also shown is the BLR size-luminosity relation \citep{ben09} implying sizes an order of magnitude or so larger than the optical disk.}
\label{lagl}
\end{figure}

We first consider the relation between the lag and the object luminosity, both of which are independent observables in our study. Figure \ref{lagl} shows the best-fit relation to the data, $\tau_c^0\propto L_{\rm opt}^{0.53\pm 0.10}$ (error bars were obtained by means of Monte Carlo simulations that take into account the errors on individual $\tau_c^0$ measurements, but do not consider the uncertainty on the luminosity). In the standard theory of accretion disks, as applies to AGN \citep{ss73,dl11}, the optical spectrum is self-similar ($L_\nu \sim \nu^{1/3}$, where $\nu$ is the frequency), hence results in $L_{\rm opt} \sim (M_{\rm BH} \dot{\mathcal{M}})^{2/3}$, where $\dot{\mathcal{M}}$ is the effective mass accretion rate (allowing for irradiation; see Appendix). The light crossing time of the optical emitting region in such a disk, $\tau_c^0\sim (M_{\rm BH} \dot{\mathcal{M}})^{1/3}$ \citep{col98}. This implies that, for a thin irradiated self-similar accretion disk one expects $\tau_c^0\sim L_{\rm opt}^{1/2}$, which agrees well with the observed relation.

One may wonder whether our $\tau_c^0$ measurements on sub-sampling times, $\tau_s$, are at all meaningful (e.g., NGC\,4051 has a formal lag of $\sim 0.2\pm0.1$ while the minimum visit interval is $\sim 0.8$\,days). The answer depends on a) the number of data points, and b) the power density spectrum of the quasar. In particular, red power density spectra, which have little power at high frequencies (e.g., $>1/\tau_c^0>1/\tau_s$), may result in robust continuum lag measurements provided enough data with good S/N exist. In fact, this justifies the use of linear interpolation schemes for evaluating correlation functions in the general context of AGN reverberation mapping \citep{cak07,kas00,ser05}. Nevertheless, sampling is known to affect time-lag measurements \citep{gr08}, and it will be of importance to repeat this experiment with better sampled data, especially for the lower luminosity objects, considering the typical values for $\tau_c^0$ determined here.

We note that while the minimum cadence of the observations is similar for all objects in the \citet{ser05} sample, the median sampling, which may be a more meaningful quantity as far as time-lag measurements are concerned, does show a weak trend with luminosity (Fig. \ref{lagl}). Nevertheless, were our time-lag measurements merely reflecting on the slightly different sampling used, $\tau_c^0$ would have shown a mere factor of $<2$ in range, in contrast to our findings. With that being said, it will be of importance to repeat this analysis for a sample of objects having a range of luminosities with luminosity-independent sampling, to verify that residual biases are indeed eliminated.

We note that Mrk\,509 is consistently an outlier in our sample; whether this is due to extreme inclination (see below) or, more likely in our opinion, due to an erroneous lag measurement (note the poor fit in Fig. \ref{tiles}), is yet to be determined.

\begin{figure}
\plotone{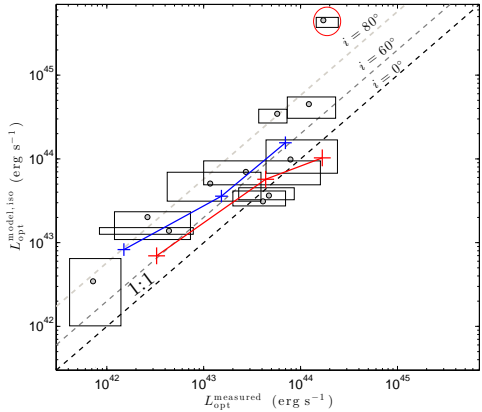}
\caption{The measured vs. predicted (isotropic) optical luminosity matches a $1:1$ relation (shown for several inclination angles), supporting the notion of geometrically-thin accretion disks being the source for the optical emission in low-luminosity type-I AGN. A notable discrepant point is that of Mrk\,509 (encircled), which could be due to erroneous lag measurement or an extremely edge-on source. The blue/red lines correspond to the geometrically-averaged results over luminosity bins (of 4/4/2 objects, from the luminous to the less luminous, and excluding Mrk\,509) assuming the average/\citet{cak07} luminosity. The results are consistent with the brighter objects in the \citet{ser05} sample being viewed face on. Results for the lowest luminosity sources may be affected by poor sampling (see \S6).}
\label{rel}
\end{figure}

\subsection{The Source of Optical Emission}

The fact that $\tau_c^0 \propto L_{\rm opt}^{1/2}$ does not necessarily imply that viscous heating (perhaps with some irradiation) is primarily responsible for the bulk of the optical emission.  For example, the light intensity of a varying point source that scatters off a uniform extended surface could result in a similar size-luminosity relation. To show that the irradiated thin accretion disk physics indeed provides a viable interpretation of the data, we will next relate the measured time-delays to the optical luminosity of the source within the framework of a \citet{ss73} accretion disk model. 

Note that, for low-amplitude radiative perturbations of a disk carrying an effective mass accretion rate, $\dot{\mathcal{M}}$ \citep[and the Appendix]{col98},
\begin{equation}
\tau_c^0\simeq 0.7 \left ( \frac{M_{\rm BH}\dot{\mathcal{M}}}{{\rm 10^6M_\odot^2~yr^{-1}}} \right )^{1/3}\,{\rm days},
\label{tau0}
\end{equation}
where the exact normalization depends on which property of the continuum transfer function is being traced (for reverberation mapping algorithms that effectively measure the centroid of the transfer function, the above normalization holds irrespectively of the disk inclination\footnote{This assumes azimuthally symmetric illuminated accretion disk.}). Our normalization of equation \ref{tau0} is smaller by a factor $\simeq 2$ than the one used by \citet[allowing for the different definitions used in each work]{col98}, and is supported by numerical calculations outlined in the Appendix. It is also in better agreement with the revised values adopted in \citet{col01}. 

The optical flux in some AGN has been known to vary significantly over long, $\gg 1$\,day, timescales \citep{ut03}, which are, nevertheless, shorter than dynamical timescales associated with the optically-emitting region. If due to variations in the irradiation level then the disk would effectively "breath" during the time-series, increasing/decreasing its size in high/low flux-states, as $\dot{\mathcal{M}}$ varies (Eq. \ref{tau0} and the Appendix). Therefore, as one extracts the disk size from the full light curves, one actually measures some average value over the time-series [c.f., the corresponding effect for the BLR \citep{pet02}]. This average is, however, difficult to define, hence to accurately interpret, since it depends on how the signal is accumulated by the correlation function and therefore on the light curve properties. Specifically, the particular variability pattern of the AGN, the sampling used, and the S/N of the data are all relevant factors in determining the exact value of the average lag. For the \citet{ser05} sample, we find that flux variations could affect the the effective disk size at the $\sim 10\%$ level for most objects (with the most pronounced effect being at the $\sim 17$\% level for 3C\,390.3). Such uncertainties should be borne in mind when quantitatively interpreting the data, and better datasets are required to shed light on their potential importance. 

Inverting equation \ref{tau0}, one can use the measured lags to determine the product $M_{\rm BH} \dot{\mathcal{M}}$, which, in turn, sets the optical luminosity for AGN \citep[and the Appendix]{bz03,col02,ld11}. This leads to the total optical disk luminosity being
\begin{equation}
L_{\rm opt}^{\rm model} \simeq 3\times 10^{43} \left ( \frac{\tau_c^0}{{\rm days}} \right )^2\,{\rm erg~s^{-1}}.
\label{lopt}
\end{equation}
To be able to compare model predictions to observations, whose reported values assume isotropic emission of the source, we note that $L_{\rm opt}^{{\rm model,iso}}=2L_{\rm opt}^{\rm model}\,{\rm cos}(i)$ where $i$ is the inclination angle. This means that, for sources observed with inclination angles $i>60^\circ$, the implied isotropic luminosity will be lower than the true luminosity of the source. Nevertheless, for type-I sources, whose continuum emission is not obscured by the putative torus, it is believed that, generally, $i<60^\circ$, which is also consistent with the findings of microlensing studies \citep{poi10}. 

Figure \ref{rel} shows model predictions for the isotropic optical luminosity ($i=0^\circ$ is assumed) vs. the measured value. It is important to re-emphasize that, a-priori, it is not clear that equation \ref{lopt} should predict a luminosity similar to the observed one as our time lag measurements are independent of the object luminosity (so long as the AGN is not too faint to be observed). The fact that a  $1:1$ relation between the measured and predicted quantities is roughly obtained (note the different $1:1$ relations plotted for various object inclinations in Fig. \ref{rel}) provides strong evidence for the prevalence of standard thin irradiated accretion disks in AGN, at least as far as their optical emission properties are concerned. Furthermore, it implies that our time lag measurements are probably in the right ball park, and that type-I AGN are not highly-inclined sources, as expected by theory. For example, had our time-lag measurements been a factor $\sim 2$ larger than observed [c.f. \citet{cak07} and our Fig. \ref{med}], the predicted luminosities would have been a factor $\sim 4$ larger, implying that AGN in the \citet{ser05} sample are either highly inclined, or that accretion disks are intrinsically inefficient emitters, in contrast with the underlying physical assumption of geometrically-thin disks \citep{ss73}. With that being said, one should be careful to not over-interpret the data, especially at the low-luminosity end, where under-sampling could lead to over-estimated lags hence a bias (see above).

\section{Discussion}

The work presented here suggests that optical light-echoes associated with the accretion flow in low-luminosity AGN is a rather general phenomenon, which may be traced not only using spectroscopic data (for which only one convincing example exists to date, NGC\,7469), but also using broadband photometric means. Nevertheless, to properly measure the time-delay associated with those echoes using photometric means, and to be able to subtract the interfering BLR signal, more elaborate techniques than cross-correlation schemes are required. 

\subsection{Accretion Disk Sizes}

Our results are consistent with the notion that the bulk of the optical luminosity in AGN is emitted by standard, geometrically-thin, irradiated accretion disks. In contrast, using data for lensed quasars, \citet{mor10} concluded that accretion flows are inconsistent with the bulk of the (rest-UV) emission originating in standard disks, as they are sub-luminous given their microlensing-implied sizes (assuming they radiate as black bodies with $T\sim r^{-3/4}$). In particular, they found that the size of their rest-UV emitting regions are on average, a factor $\sim 4$ larger than predicted by theory for plausible values of the radiative efficiency, object inclination, and black hole mass. Crudely, had the same effects been present here, our predicted luminosities would have been $\sim 16$ times greater than observed (Eq. \ref{lopt}), which is, generally, not the case\footnote{For such a discrepancy to be accommodated by our rest-frame optical data would require that our time-lags measurements are systematically under-estimated and/or our luminosities systematically over-estimated, and that all objects are, in fact, lying on a relation whose normalization is set by our brightest object, Mrk\,509; see figure \ref{rel}.}. Whether this indicates a departure from self-similarity in the inner disk regions of AGN, a genuine difference between quasars and AGN, or systematic effects (and/or inadequate interpretation of the signal) in either of the methods, requires further investigation. In what follows, we shall {\it assume} that geometrically-thin accretion disks provide a viable physical explanation for the bulk of the optical emission in AGN,  and consider the physical implications of our results.

\begin{figure*}
\plottwo{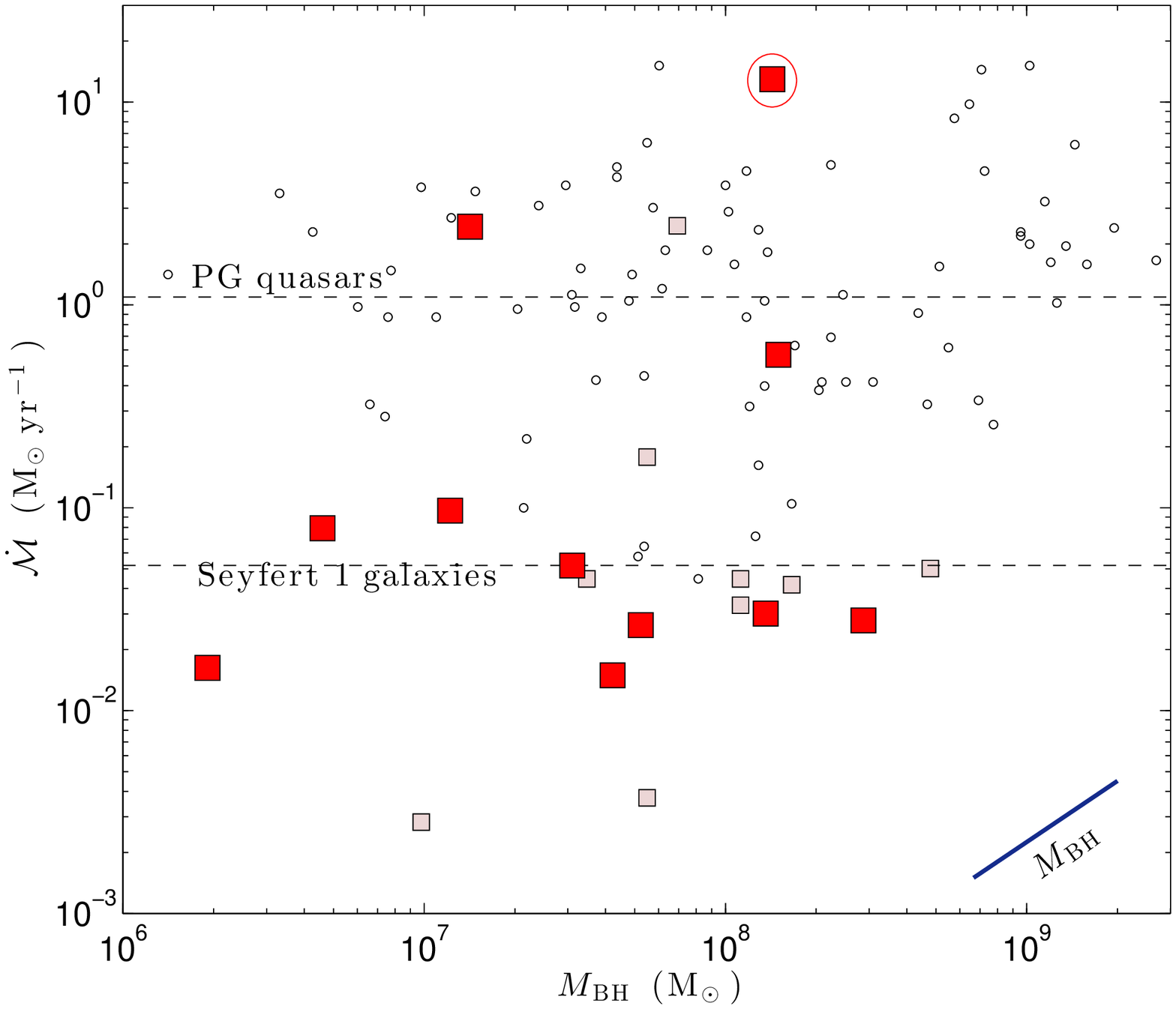}{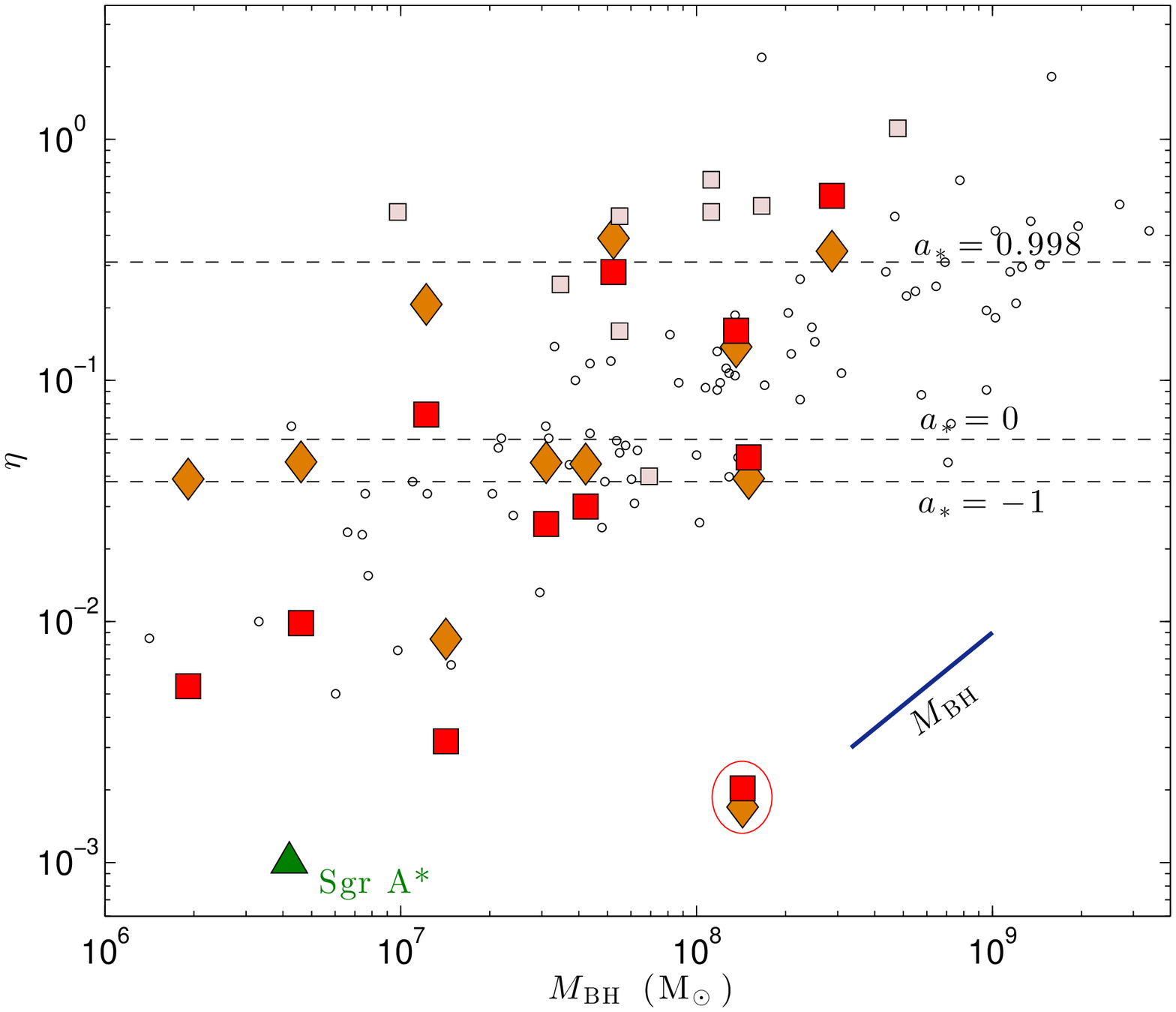}
\caption{Effective mass accretion rates, $\dot{\mathcal{M}}$ (left panel), and radiative efficiencies, $\eta$, (right panel) for accretion flows in AGN as a function of the reverberation-based black hole mass, $M_{\rm BH}$. {\it Left:}  Comparing $\dot{\mathcal{M}}(M_{\rm BH})$ estimates to those of the PG quasars \citet[black points]{dl11}, we find that Seyfert galaxies accrete at an order of magnitude or more lower rates. $\dot{\mathcal{M}}$-values for most objects are similar, and comparable to those deduced by \citet{r12} based on independent optical luminosity considerations. {\it Right:} the accretion efficiency, $\eta(M_{\rm BH})$, assuming that the bolometric luminosity is (a) $9\times\lambda L_\lambda(5100\,{\rm \AA})$ \citep[red squares; see also Table 1]{dl11} or that (b) it inversely scales with mass \citep[orange diamonds; see \S6.2]{r12}. Under the same set of assumptions, results for the PG quasars are similar to those of Seyfert galaxies, and there is tentative indication for $\eta$ being a rising function of $M_{\rm BH}$. Over-plotted are efficiency predictions from \citet{nt73} where $a_*$ denotes the BH spin, and the value deduced for Sgr\,A$^*$ \citep{qg00}. The grey line shows the effect of a 0.5\,dex shift in mass on the results. Changing the bolometric corrections would shift objects along the ordinate. Results for Mrk\,509 are encircled (see also Fig. \ref{rel}).}
\label{mdot}
\end{figure*}

\subsection{Effective Mass accretion rates and radiative efficiencies}

Our time lag measurements, as interpreted by standard thin (infinite) accretion disk models, may be used to measure the effective mass accretion rate, $\dot{\mathcal{M}}$, {\it independently} of the luminosity, in objects for which $M_{\rm BH}$ is known (Eq. \ref{tau0})\footnote{While the results of the previous section imply that whether one uses $\tau_c^0$ to measure $\dot{\mathcal{M}}$ or the optical luminosity, is immaterial, using time lag measurements has the advantage of being less affected by systematics such as extinction and reddening in the AGN host, and to the latter's contribution to the flux (this may be a non-negligible source for uncertainty given the relatively large apertures used by \citet{ser05}). Also, mass accretion rates estimated from the centroid lags are unaffected by accretion disk inclination, in contrast to luminosity-based estimates \citep{dl11,r12}.}. Typical  $\dot{\mathcal{M}}$ values for objects in the \citet{ser05} sample are shown in figure \ref{mdot}, where the results hover around $0.01$-$0.1\,{\rm M_\odot~yr^{-1}}$. When contrasted with the accretion rates of the Palomar-Green objects \citep{dl11}, Seyfert 1 galaxies tend to have $\dot{\mathcal{M}}$ values that are smaller by about $1$-$2$ orders of magnitude, which is  consistent with their luminosities being similarly lower than bona-fide quasars. Comparing our results for the mass accretion rate to those of other Seyfert 1 galaxies from \citet{r12}, based on optical luminosity considerations \citep{bec87}, we find qualitative agreement with their findings. We note, however, that the true mass accretion rate is somewhat lower than deduced here, as $\dot{\mathcal{M}}$ includes the effects of irradiation; this is also true for the \citet[see also \citet{r12}]{dl11} results (see Appendix).

An interesting property of our results is that the deduced $\dot{\mathcal{M}}$ values (apart from three objects, including Mrk\,509) span only about a factor of $\sim 5$ in range, and could even be similar given the measurement uncertainties (mainly on $M_{\rm BH}$). In contrast, the AGN in our sample span almost two orders of magnitude in luminosity. Whether this implies a common triggering mechanism for low-luminosity type-I AGN, which sets a rather uniform mass accretion rate across a broad range of BH masses, or whether this result vanishes as better data and more statistics are accumulated, remains to be determined. It is, however, interesting to note that a similar non-dependence of the (median) mass accretion rate on $M_{\rm BH}$ has been previously reported by \citet[see our Fig. \ref{mdot}]{dl11}, for the PG quasars, over a similar range of black hole masses.

It is furthermore possible to estimate the radiative efficiency of Seyfert galaxies,  $\eta \equiv L_{\rm bol}/\dot{\mathcal{M}}c^2$, provided the bolometric luminosity of the source, $L_{\rm bol}$ may be reasonably estimated from the limited spectral range probed. To this end, we first assume that $L_{\rm bol}/L_{\rm opt}\simeq 9$, and is independent of $M_{\rm BH}$, as was assumed for the PG sample of quasars \citep{dl11}. Alternatively, we assume that $L_{\rm bol}/L_{\rm opt}\sim 9(M_{\rm BH}/10^8\,{\rm M_\odot})^{-0.5}$, which is consistent with the findings of \citet[see also \citet{sc04}]{r12} for their sample of AGN. We plot $\eta(M_{\rm BH})$ in figure \ref{mdot} using the above bolometric corrections, and conclude that the radiative efficiency is indeed consistent with the typical efficiencies from geometrically-thin accretion disk models \citep{nq05}. The results for a constant bolometric correction trace well the findings of \citet{dl11} for the PG sample of quasars. This means that, under the same assumptions, the radiative efficiency, $\eta$, for PG quasars and for Seyfert 1 galaxies behaves in a similar manner with the BH mass, and is independent of the mass accretion rate\footnote{Note that both the method applied here, and the one used by \citet{dl11}, measure $\dot{\mathcal{M}}$ rather than $\dot{M}$; see the Appendix.}. Using a mass-dependent bolometric correction does not appreciably alter the above conclusions. 

Our findings provide tentative evidence for $\eta$ being a rising function of black hole mass, as previously reported by \citet{dl11} based on SED arguments. With that being said, better bolometric corrections for individual objects, as well as better time-delay measurements for the least massive AGN are required to firmly establish such a trend, and, if real, its origin.

\subsection{Reverberation mapping of accretion disks}

In this study we were able to measure, for the first time in a non-biased way and with a high success-rate, continuum-to-continuum time-delays in 11/12 low-luminosity AGN. To securely determine the time-delays associated with an irradiated accretion disk, high cadence, good S/N observations are required, and prior knowledge of the BLR size could be advantageous in narrowing down the parameter-space, thereby leading to higher-confidence results. 

Evidently, minimal benchmarks for a successful reverberation campaign are provided by the \citet{ser05} sample. Nevertheless, given the shorter delays deduced here compared to the \citet{ser05} findings, higher cadence observations, especially for low-luminosity objects are preferred. With upcoming photometric surveys, such as the {\it Large Synoptic Survey Telescope} (especially their deep-drilling fields), it may be possible to obtain much higher quality data for many more (bright) objects, potentially transforming the field. 

In the long term, it may be possible to look for deviations from simple infinite self-similar disk models, which could shed light on the role of outflows in those systems, as well as on the accretion rate in the inner disk regions, which ultimately feeds the BH. Further, it may be possible to probe the far outskirts of disk, where it becomes self-gravitating, and may join with other components of the AGN, such as the BLR. In addition, it may be possible to determine the inclinations of accretion disks by comparing their apparent luminosities to their predicted ones. Lastly, reverberation mapping of the accretion flows in numerous AGN could reveal the different modes of accretion prevailing in those systems (e.g., slim vs. thin disk accretion) with implications for black hole growth rates.  

Once accretion disk models are quantitatively tested and verified, and reddening corrections, and the host contribution to the flux adequately estimated, it may be possible to use quasars as standard candles, as envisioned by \citet[and references therein]{col01,cak07}. Interestingly, as our deduced lags are a factor $\sim 2$ smaller than those considered by \citet{cak07}, with all other quantities held fixed at the values adopted by these authors, our deduced value for $H_0$ based on their formalism (see their Eq. 15) would be in the range $42-88\,{\rm km~s^{-1}Mpc^{-1}}$ and in better agreement with concordance cosmology. The full treatment of this topic is beyond the scope of this paper.

\section{Summary}

Analyzing the photometric light curves for 12 objects in five bands from the \citet{ser05} sample we show that light echoes propagating across the continuum emission region is a general phenomenon in AGN and, provided that appropriate reverberation mapping techniques are employed, can be used to reliably infer the size of accretion disks, and probe their physics. We further show that the results are consistent with time-delays that originate in geometrically-thin, radiatively-efficient accretion disks, which provide independent support of the standard paradigm of BH accretion in low-luminosity type-I AGN.

Interpreting the measured lags in the framework of irradiated accretion disk models, we are able to determine, using measurable quantities independent of the optical flux, the mass accretion rate in Seyfert 1 galaxies. The typical mass accretion rates are found to be $\lesssim 0.1 \,{\rm M_\odot~yr^{-1}}$ over the full black hole mass range explored here, and about an order of magnitude lower than those that characterize the PG quasars \citep{dl11}. We find independent, yet tentative, support for the findings of \citet{dl11}, where the radiative efficiency of AGN, $\eta$, increases with their BH mass, and show that $\eta$ is, at least to zeroth-order effect, independent of the mass accretion rate (hence luminosity) of these sources. 

The upshot of this work is that simple accretion disks seem to provide an adequate description for the physics underlying the optical emission in AGN, and that whether one uses time-delays or optical emission to infer their mass accretion rates should lead to similar results. With large upcoming photometric surveys that will monitor the sky with high-cadence, in several broadband filters, it may be possible to measure continuum time-delays, hence mass accretion rates and intrinsic (rest optical) luminosities in numerous objects, in a way which is independent of inclination, host contamination, and reddening effects. In the coming decade we are therefore likely to experience a significant progress in our understanding of accretion phenomena in AGN with implications for BH growth and feedback processes.

\acknowledgements 

We are indebted to S. G. Sergeev and collaborators for acquiring an exquisite photometric dataset for AGN that made this work possible. We are especially grateful to A. Laor and H. Netzer for carefully reading an earlier version of this paper, and for insightful comments, as well as an anonymous referee for important feedback. We thank M. Gaskell for a motivating email exchange, and S. Kaspi, C. Montuori, and S. Rafter for fruitful discussions. This research has been supported in part by a FP7/IRG PIRG-GA-2009-256434 and by grant 927/11 from the Israeli Science Foundation and the Jack Adler Foundation.

\appendix \section{Irradiated Accretion Disk Models}

We wish to interpret time-lag measurements within the framework of the irradiated, self-similar and geometrically-thin, accretion disk model (i.e., to determine the normalization of Eq. \ref{tau0}) so that the mass accretion rate may be estimated. To this end, we resort to semi-analytic models and consider a \citet{ss73} accretion disk solution around a BH of mass $M_{\rm BH}$ with a mass accretion rate $\dot{M}$. We solve for the emitted spectral energy distribution assuming local dissipation of gravitational energy, and black body emission from each annulus in the disk. We neglect general relativistic effects near the BH, which is justified in the optical band. 

Consider next a case where the above disk is further illuminated by central extreme-UV (EUV) point sources located at an elevation $h_{\rm EUV}$ above the disk surfaces\footnote{The source need not be elevated above the disk and can, in fact, be part of the inner disk. For example, if the effective disk thickness is $\propto \alpha r$ ($\alpha$ being the viscosity parameter) then, as recent magnetohydrodynamic simulations suggest \citep[and references therein]{pen12}, the rise of $\alpha$ near the black hole may provide a natural elevated continuum source illuminating the outer disk. Also, a modest flaring of the outer disk could make the inner region irradiate its outskirts. Lastly, a compact (ionized) reflector covering a fair fraction of the sky around the inner accretion disk may also result in illumination of the outer disk parts.}. For an EUV source luminosity given by $L_{\rm EUV}(t)$, the local irradiated disk temperature, assuming instantaneous reprocessing, is set by \citep[and references therein]{cak07},
\begin{equation}
T(r,t)= \left [ \frac{3GM_{\rm BH}\dot{\mathcal{M}}(t-r/c)}{8\pi \sigma r^3} \right ]^{1/4},
\label{tr}
\end{equation}
where the effective mass accretion rate,
\begin{equation}
\dot{\mathcal{M}}(t) \equiv \dot{M}\left [1+\frac{8\pi}{3} \left ( \frac{\dot{m}}{\dot{M}} \right )  \frac{L_{\rm EUV} (t)}{L_{\rm Edd.}} \right ] \geq \dot{M},
\label{dotm}
\end{equation}
and $\dot{m}\equiv m_p (c/r_g) (h_{\rm EUV}/r_g) (r_g^2/\sigma_T)$ ($r_g\equiv GM_{\rm BH}/c^2$, $\sigma_T$ is the Thomson cross-section, $L_{\rm Edd.}$ is the Eddington luminosity, and $m_p$ the proton mass).  Interestingly, for slow perturbations (i.e., with appreciable flux variations over timescales longer than the light-crossing time-scale of the disk), the optical luminosity satisfies, $L_{\rm opt}\propto \dot{\mathcal{M}}^{2/3}\propto ( L_{\rm EUV}+{\rm const.})^{2/3}$ (\S6.2), which could explain the fact that quasars become bluer as they brighten \citep{sh12}, provided that some residual EUV emission leaks into the near-UV band.

To put some numbers, for a Seyfert 1 galaxy emitting at $0.1L_{\rm Edd.}$, we shall assume $L_{\rm EUV}/L_{\rm Edd}= 0.03$. Now, EUV emission originates close to the black hole, at  $<10^2r_g$, and so, for a thin accretion disk, where the outer-disk viscosity parameter, $\alpha\sim 0.01$ \citep{pen12}, it is reasonable to assume that $h_{\rm EUV}\sim r_g$. Taking $M_{\rm BH}\sim 10^8\,{\rm M_\odot}$ and $\dot{M}\sim 10^{-2}\,{\rm M_\odot~yr^{-1}}$, we get $\dot{\mathcal{M}}/\dot{M} \sim 1.5$. As the optical luminosity, $L_{\rm opt} \propto \dot{\mathcal{M}}^{2/3}$, typical brightening levels over timescales greater than the light crossing times of the disk can reach the 30\% level, which is of the order of the observed values in AGN (note that we do not claim here that irradiation is the only cause for optical AGN variability, on all timescales). Further, as $\dot{\mathcal{M}}/\dot{M}-1 \propto L_{\rm EUV}$, and irradiation may have a constant flux component, as well as a varying one, our ability to accurately measure the mass accretion rate, $\dot{M}$, is limited, as a constant irradiation level can always masquerade as a higher effective accretion rate, $\dot{\mathcal{M}}$. However, if accretion rather than irradiation is primarily responsible for the optical emission in AGN then $\dot{\mathcal{M}} \gtrsim \dot{M}$, and irradiation may be considered as a small perturbation.

\begin{figure}
\plottwo{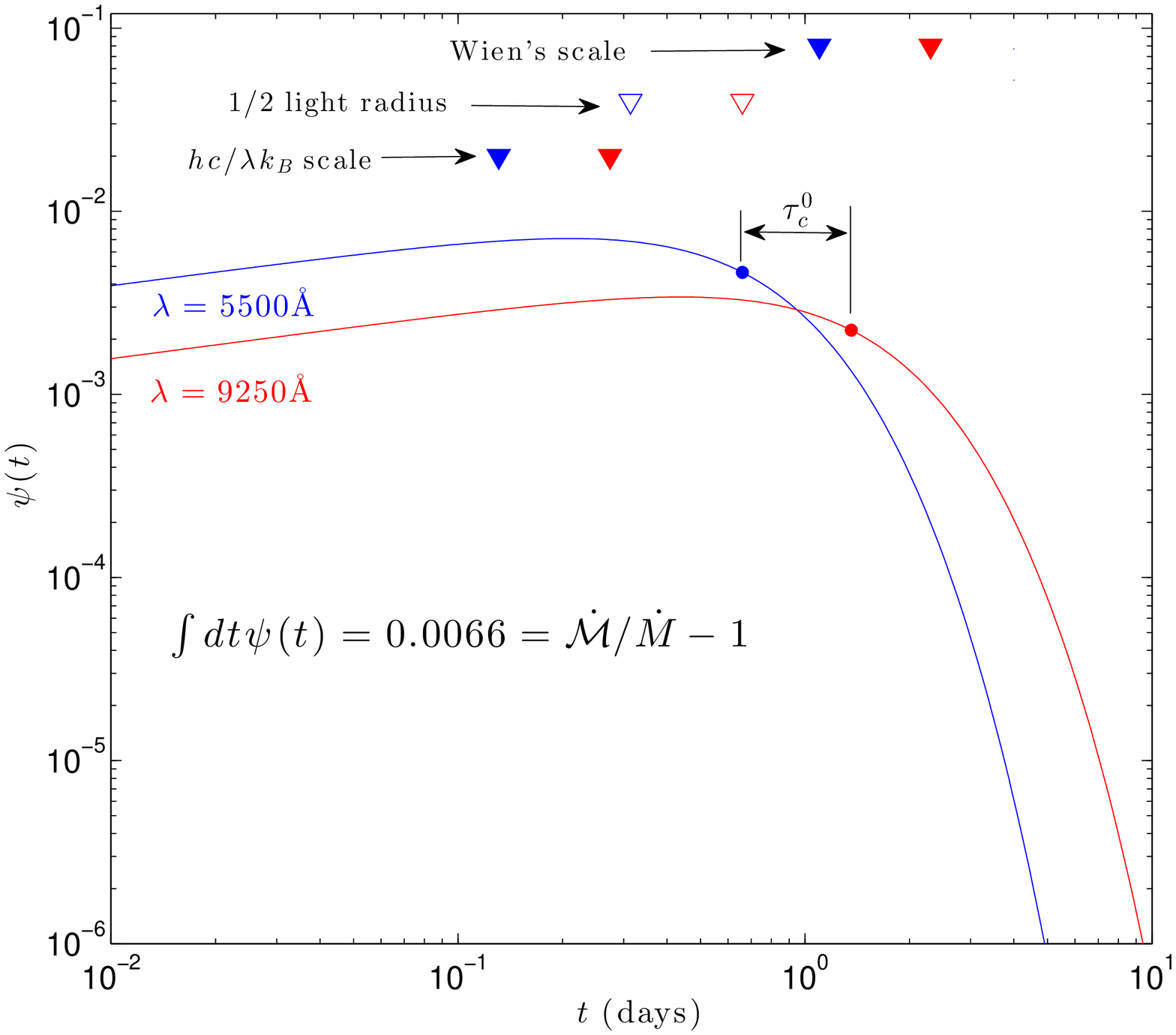}{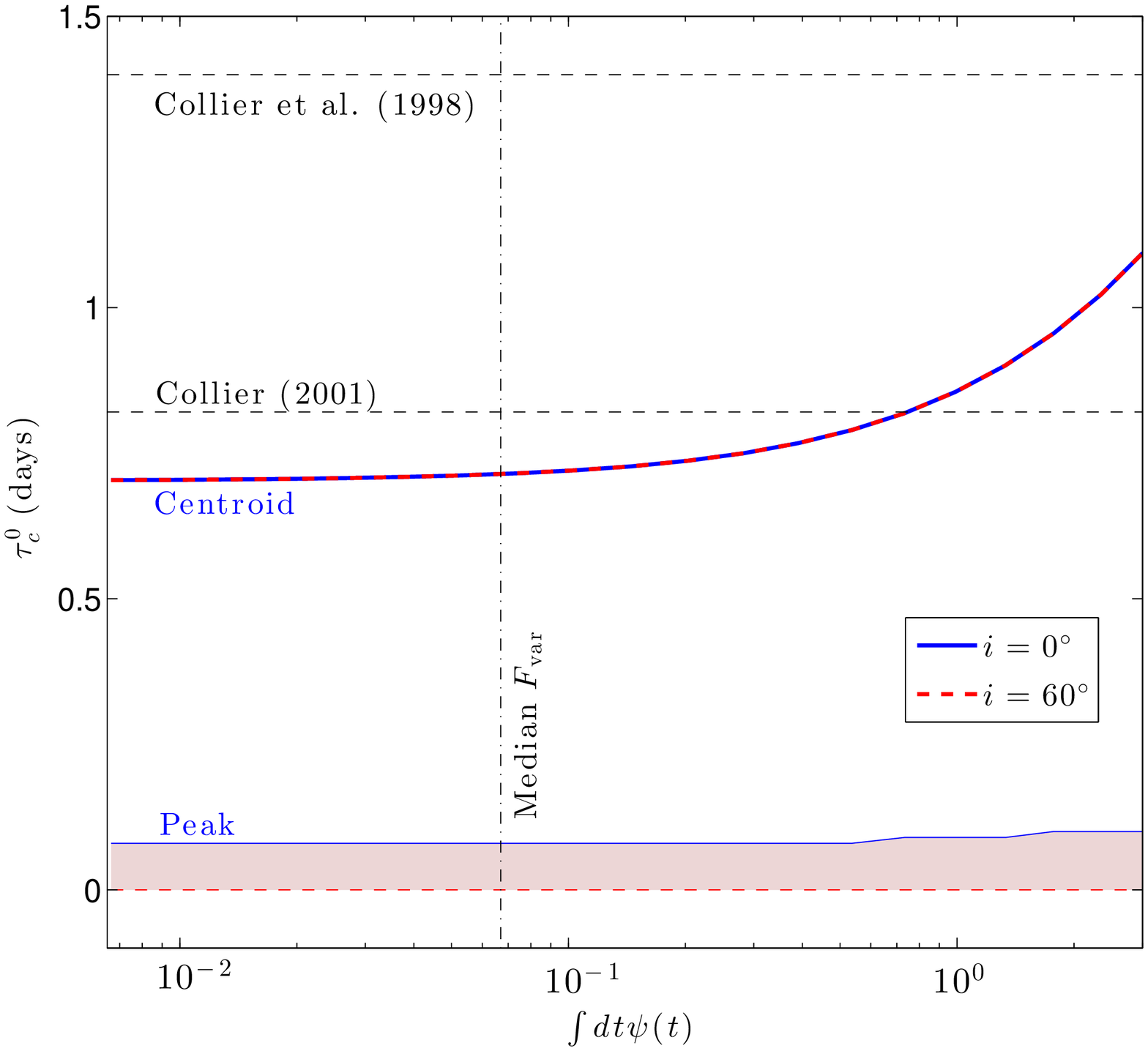}
\caption{The propagation of perturbations across a standard, self-similar, accretion disk viewed face on, and characterized by $M_{\rm BH}\dot{M}=10^6\,{\rm M_\odot^2~yr^{-1}}$. {\it Left:} The transfer function at 5500\AA\ (blue) and 9250\AA\ (red) for a  accretion disk subjected to a perturbation of a $\delta$-function form from a point source at its center. The transfer functions are normalized such that  $\int dt \psi(t)=\dot{\mathcal{M}}/\dot{M}-1 \propto L_{\rm EUV}$ (i.e., had $L_{\rm EUV}$ been a constant, the disk would appear to have an effective mass accretion rate, $\dot{\mathcal{M}}$). The centroids of the transfer functions are denoted by filled circles. For comparison we also show the three additional wavelength-dependent scales: the half-light radius, and $r(T(\lambda))$, with $T(\lambda)$ either obtained from Wien's law or by requiring that  $T=hc/\lambda k_B$. {\it Right:} The relation between $\tau_c^0$ (as would be measured from the centroid or the peak of the transfer functions, for two disk inclinations; see legend) and the amplitude of the perturbation. For low amplitude perturbations, $\tau_c^0$ is constant as the disk temperature structure does not deviate considerably from its non-perturbed state. For higher-amplitude perturbations, the optically emitting region increases in size, and the measured lag corresponds to its size had the perturbation acted for times longer than its light crossing time. Specifically, the curve corresponding to the centroid lag follows a $\tau_c^0\sim [1+\int dt \psi(t)]^{1/3}$ relation. For low-level perturbations ($\sim 7$\%, i.e., of order the median fractional variability for the sample; see dash-dotted line), with lag measurements that trace the centroid of the of the transfer function, $\tau_c^0\simeq 0.7$\,days (c.f. Eq. \ref{tau0}), which is below the value adopted in \citet{col98}, and closer to the value considered in \citet{col01}, and is independent of the disk inclination.}
\label{trans}
\end{figure}

To calculate the normalization of equation \ref{tau0}, we evaluate the continuum transfer functions, $\psi(t)$, at 5500\AA\ and at 9250\AA\ (the time-delay between continuum light curves in those two bands corresponds to $\tau_c^0$), for a short pulse, with respect to the light crossing time of the disk, of irradiating flux that peaks at a luminosity $L_{\rm EUV}$, and show the results in figure \ref{trans}. Clearly, higher levels of irradiating flux result in more considerable departures from the non-irradiated case, where $\dot{\mathcal{M}}/\dot{M}-1=0$.  The centroid time-delay (i.e., $\tau_c^0$) between the bands is given by the centroid of $\psi=\psi_{\rm 5500\AA }*\psi_{\rm 9250\AA }$ ('$*$' denotes convolution). Results for a non-irradiated disk with $M_{\rm BH}\dot{M}=10^6\,{\rm M_\odot^2~yr^{-1}}$ are shown in figure \ref{trans} for various levels of irradiation. As expected, for low levels of irradiation ($\dot{\mathcal{M}}/\dot{M}-1< 0.1$),  $\tau_c^0\simeq 0.7 (M_{\rm BH}\dot{M}/10^6\,{\rm M_\odot^2~yr^{-1}})^{1/3}$ to a good approximation (see \S5.2 and Eq. \ref{tau0}). For higher levels of irradiation, the measured disk size reflects on the irradiated structure rather than on the non-perturbed state, hence, asymptotically, $\tau_c^0\propto (M_{\rm BH} L_{\rm EUV})^{1/3}$ (see the right panel of Fig. \ref{trans} and its caption, as well as \S5.2).

Note that, on accounts of our assumption of azimuthal symmetry, inclination does not affect the centroid lag. For comparison, we also show the time-difference between the peaks of the correlation functions in each band, which is shifted to smaller lags, and is sensitive to inclination effects.  


\begin{thebibliography}{}

\bibitem[Abramowicz et al.(1988)]{ab88} Abramowicz, M.~A., Czerny, B., Lasota, J.~P., \& Szuszkiewicz, E.\ 1988, \apj, 332, 646

\bibitem[Bachev(2009)]{bac09} Bachev, R.~S.\ 2009, \aap, 493, 907

\bibitem[Barth et al.(2013)]{bar13} Barth, A.~J., Pancoast, A., Bennert, V.~N., et al.\ 2013, arXiv:1304.4643

\bibitem[Bechtold et al.(1987)]{bec87} Bechtold, J., Czerny, B., Elvis, M., Fabbiano, G., \& Green, R.~F.\ 1987, \apj, 314, 699

\bibitem[Bentz et al.(2006)]{ben06} Bentz, M.~C., Peterson, B.~M., Pogge, R.~W., Vestergaard, M., \& Onken, C.~A.\ 2006, \apj, 644, 133

\bibitem[Bentz et al.(2009)]{ben09} Bentz, M.~C., Peterson, B.~M., Netzer, H., Pogge, R.~W., \& Vestergaard, M.\ 2009, \apj, 697, 160

\bibitem[Bentz et al.(2010)]{ben10} Bentz, M.~C., Walsh, J.~L., Barth, A.~J., et al.\ 2010, \apj, 716, 993

\bibitem[Bian et al.(2010)]{bi10} Bian, W.-H., Huang, K., Hu, C., et al.\ 2010, \apj, 718, 460

\bibitem[Bian \& Zhao(2003)]{bz03} Bian, W.-H., \& Zhao, Y.-H.\ 2003, \pasj, 55, 599

\bibitem[Blandford \& Begelman(1999)]{bb99} Blandford, R.~D., \& Begelman, M.~C.\ 1999, \mnras, 303, L1 


\bibitem[Blackburne et al.(2011)]{bla11} Blackburne, J.~A., Pooley, D., Rappaport, S., \& Schechter, P.~L.\ 2011, \apj, 729, 34

\bibitem[Cackett et al.(2007)]{cak07} Cackett, E.~M., Horne, K., \& Winkler, H.\ 2007, \mnras, 380, 669 

\bibitem[Chelouche \& Daniel(2012)]{cd11} Chelouche, D., \& Daniel E.\ 2012, \apj, 747, 62 

\bibitem[Chelouche et al.(2012)]{cd12} Chelouche, D., Daniel, E., \& Kaspi, S.\ 2012, \apjl, 750, L43

\bibitem[Chelouche \& Zucker(2013)]{cz12} Chelouche, D., \& Zucker S.,\ 2013, \apj, in press 


\bibitem[Collier et al.(1998)]{col98} Collier, S.~J., Horne, K., Kaspi, S., et al.\ 1998, \apj, 500, 162

\bibitem[Collier(2001)]{col01} Collier, S.\ 2001, \mnras, 325, 1527

\bibitem[Collier et al.(2001)]{col01b} Collier, S., Crenshaw, D.~M., Peterson, B.~M., et al.\ 2001, \apj, 561, 146

\bibitem[Collin et al.(2002)]{col02} Collin, S., Boisson, C., Mouchet, M., et al.\ 2002, \aap, 388, 771

\bibitem[Davis \& Laor(2011)]{dl11} Davis, S.~W., \& Laor, A.\ 2011, \apj, 728, 98

\bibitem[Dexter \& Agol(2011)]{da11} Dexter, J., \& Agol, E.\ 2011, \apjl, 727, L24

\bibitem[Edri et al.(2012)]{ed12} Edri, H., Rafter, S.~E., Chelouche, D., Kaspi, S., \& Behar, E.\ 2012, \apj, 756, 73

\bibitem[Glikman et al.(2006)]{gl06} Glikman, E., Helfand, D.~J., \& White, R.~L.\ 2006, \apj, 640, 579

\bibitem[Grier et al.(2008)]{gr08} Grier, C.~J., Peterson, B.~M., Bentz, M.~C., et al.\ 2008, \apj, 688, 837

\bibitem[Grier et al.(2012)]{gr12} Grier, C.~J., Peterson, B.~M., Pogge, R.~W., et al.\ 2012, \apj, 755, 60

%\bibitem[Gultekin \& Miller(2012)]{gm12} Gultekin, K., \& Miller, J.~M.\ 2012, arXiv:1207.0296

\bibitem[Hu et al.(2008)]{hu08} Hu, C., Wang, J.-M., Ho, L.~C., et al.\ 2008, \apj, 687, 78 

\bibitem[Kaspi et al.(2000)]{kas00} Kaspi, S., Smith, P.~S., Netzer, H., Maoz, D., Jannuzi, B.~T., \& Giveon, U.\ 2000, \apj, 533, 631

\bibitem[Laor \& Davis(2011)]{ld11} Laor, A., \& Davis, S.~W.\ 2011, \mnras, 417, 681

\bibitem[Laor \& Netzer(1989)]{ln89} Laor, A., \& Netzer, H.\ 1989, \mnras, 238, 897

\bibitem[Lynden-Bell(1969)]{lyn69} Lynden-Bell, D.\ 1969, \nat, 223, 690

\bibitem[Malkan(1983)]{mal83} Malkan, M.~A.\ 1983, \apj, 268, 582

\bibitem[Morgan et al.(2010)]{mor10} Morgan, C.~W., Kochanek, C.~S., Morgan, N.~D., \& Falco, E.~E.\ 2010, \apj, 712, 1129

\bibitem[Narayan \& Quataert(2005)]{nq05} Narayan, R., \& Quataert, E.\ 2005, Science, 307, 77 

\bibitem[Narayan \& Yi(1994)]{ny94} Narayan, R., \& Yi, I.\ 1994, \apjl, 428, L13 

\bibitem[Noble et al.(2011)]{nob11} Noble, S.~C., Krolik,  J.~H., Schnittman, J.~D., \& Hawley, J.~F.\ 2011, \apj, 743, 115

\bibitem[Novikov \& Thorne(1973)]{nt73} Novikov, I.~D., \& Thorne, K.~S.\ 1973, Black Holes (Les Astres Occlus), 343

\bibitem[Penna et al.(2012)]{pen12} Penna, R.~F., Sadowski, A., Kulkarni, A.~K., \& Narayan, R.\ 2012, arXiv:1211.0526

\bibitem[Peterson et al.(2002)]{pet02} Peterson, B.~M., Berlind, P., Bertram, R., et al.\ 2002, \apj, 581, 197

\bibitem[Peterson et al.(2004)]{pet04} Peterson, B.~M., et al.\ 2004, \apj, 613, 682

\bibitem[Poindexter et al.(2008)]{poi08} Poindexter, S., Morgan, N., \& Kochanek, C.~S.\ 2008, \apj, 673, 34

\bibitem[Poindexter \& Kochanek(2010)]{poi10} Poindexter, S., \& Kochanek, C.~S.\ 2010, \apj, 712, 668

\bibitem[Quataert \& Gruzinov(2000)]{qg00} Quataert, E., \& Gruzinov, A.\ 2000, \apj, 545, 842

\bibitem[Rafter et al.(2013)]{raf13} Rafter, S.~B., Kaspi, S., Chelouche, D., et al.\ 2013, \apj, submitted

\bibitem[Raimundo et al.(2012)]{r12} Raimundo, S.~I., Fabian, A.~C., Vasudevan, R.~V., Gandhi, P.,  \& Wu, J.\ 2012, \mnras, 419, 2529

\bibitem[Salpeter(1964)]{sal64} Salpeter, E.~E.\ 1964, \apj, 140, 796

\bibitem[Scott et al.(2004)]{sc04} Scott, J.~E., Kriss, G.~A., Brotherton, M., et al.\ 2004, \apj, 615, 135

\bibitem[Sergeev et al.(2005)]{ser05} Sergeev, S.~G., Doroshenko, V.~T., Golubinskiy, Y.~V., Merkulova, N.~I.,  \& Sergeeva, E.~A.\ 2005, \apj, 622, 129

\bibitem[Shakura(1972)]{sh72} Shakura, N.~I.\ 1972, \azh, 49, 921

\bibitem[Shakura \& Sunyaev(1973)]{ss73} Shakura, N.~I., \& Sunyaev, R.~A.\ 1973, \aap, 24, 337

\bibitem[Schmidt et al.(2012)]{sh12} Schmidt, K.~B., Rix, H.-W., Shields, J.~C., et al.\ 2012, \apj, 744, 147

%\bibitem[Slone \& Netzer(2012)]{sn12} Slone, O., \& Netzer, H.\ 2012, \mnras, 426, 656 

\bibitem[Uttley et al.(2003)]{ut03} Uttley, P., Edelson, R., McHardy, I.~M., Peterson, B.~M., \& Markowitz, A.\ 2003, \apjl, 584, L53

\bibitem[Zel'dovich(1964)]{zel64} Zel'dovich, Y.~B.\ 1964, Soviet Physics Doklady, 9, 195

\end{thebibliography}
\end{document}